# Preferential orientation of NV defects in CVD diamond films grown on (113)-oriented substrates

M. Lesik[1], T. Plays[1], A. Tallaire[2*], J. Achard[2], O. Brinza[2], L. William[2], M. Chipaux[3#], L. Toraille[3], T. Debuisschert[3], A. Gicquel[2], J.F. Roch[1] and V. Jacques[1**]

[1] *Laboratoire Aimé Cotton, CNRS, Université Paris-Sud and ENS Cachan, 91405 Orsay, France*

[2] *Laboratoire des Sciences des Procédés et des Matériaux (LSPM), Université Paris 13, Sorbonne Paris Cité, CNRS, 93430 Villetaneuse, France*

[3] *Thales Research & Technology, 1 avenue Augustin Fresnel, 91767 Palaiseau Cedex, France*

**Abstract**

Thick CVD diamond layers were successfully grown on (113)-oriented substrates. They exhibited smooth surface morphologies and a crystalline quality comparable to (100) electronic grade material, and much better than (111)-grown layers. High growth rates (15-50 µm/h) were obtained while nitrogen doping could be achieved in a fairly wide range without seriously imparting crystalline quality. Electron spin resonance measurements were carried out to determine NV centers orientation and concluded that one specific orientation has an occurrence probability of 73 % when (100)-grown layers show an equal distribution in the 4 possible directions. A spin coherence time of around 270 µs was measured which is equivalent to that reported for material with similar isotopic purity. Although a higher degree of preferential orientation was achieved with (111)-grown layers (almost 100 %), the ease of growth and post-processing of the (113) orientation make it a potentially useful material for magnetometry or other quantum mechanical applications.

\# Currently at University of Groningen / University Medical Center Groningen, Department of Biomedical Engineering, Antonius Deusinglaan 1, 9713 AV Groningen, The Netherlands

Corresponding authors:
\* alexandre.tallaire@lspm.cnrs.fr
\*\* vjacques@ens-cachan.fr

## 1. Introduction

The negatively charged nitrogen-vacancy center (NV) in diamond has focused a lot of attention in the past few years due to a number of emerging applications in quantum information [1] and magnetic sensing [2], for which it is believed to bring a substantial advantage over incumbent technologies [3]. The electronic spin state associated with this point-like defect can be optically detected [4], and coherently manipulated using microwave fields with long coherence times, even at room temperature [5]. Under an external magnetic field, splitting of the NV center's spin state occurs which can be detected by applying an appropriate resonant microwave field. Based on this principle, promising magnetic sensors achieving a high spatial resolution [6] and an extremely high sensitivity in the sub-picotesla range [7] have been reported.

For this application, the environment of the NV center in the diamond matrix needs to be carefully controlled since the close proximity of other defects can dramatically shorten coherence times [8]. Ultra-pure Chemically Vapor Deposited (CVD) single crystal diamond films in which nitrogen atoms are implanted and annealed are commonly used but this process suffers from a relatively low yield and co-implanted defects that cannot be completely annealed out [9,10]. On the other hand, grown-in NV centers that are produced by adding nitrogen during CVD diamond synthesis [11] have led to the longest coherence times so far [5]. Although spatially localizing NV centers is particularly challenging [12], an additional advantage of this approach comes from the ability to control the NV defect orientation in the diamond matrix.

Due to its $C_{3v}$ symmetry, the NV center can be oriented along one of the 4 different {111} crystallographic axes. In most diamond samples, these orientations are occupied with equal probabilities, leading to significant limitations for quantum information and sensing applications. Partial preferential orientation (50 %) of grown-in centers has been obtained at the inclined step facets of a (100) crystal grown under step-flow mode or by using a (110) substrate for growth [13,14]. More recently, an almost perfect alignment for NV centers (97-100 %) has also been achieved for (111)-grown CVD layers [15,16,17]. In this latter case, the orientation of the center perpendicularly to the surface is ideal for coupling to photonic structures and achieving enhanced light collection efficiency, potentially resulting in more robust and sensitive sensors [18].

However, the synthesis of nitrogen-doped CVD diamond layers on (111) substrates is plagued by the formation of twins and extended defects [19,20]. The extremely narrow range of optimal growth conditions required [21,22] leads to low growth rates (a few μm/h at most) and make it difficult to tune NV density in the layers. Moreover high residual stress can be responsible for crack formation and strong background blue luminescence. Post-growth processing to prepare freestanding plates or membranes is delicate to achieve due to the difficulty in cutting and polishing on this orientation [23]. Finally, (111) substrates sliced from larger crystals made by HPHT [24] or CVD are poorly available and only in small size of a few mm².

(113) crystal orientation, which is stable under certain CVD growth conditions [25], could be a potential substitute for (111) substrates. This stability is likely related to the fact that the (113) plane undergoes a surface reconstruction in the presence of a hydrogen plasma in a similar fashion to what has been reported for silicon, thus decreasing the surface energy on this orientation [26,27]. The (113) plane can be obtained by polishing a standard (100) surface with a 25.2° angle towards a <111> direction. Therefore it lies in between the (100) and (111) planes. Silva et al. [28] have considered this orientation in a geometrical model aiming at describing the evolution of crystal shape and have shown that they could allow obtaining enlarged crystals.

In this work, we present the successful growth of thick CVD layers on (113)-oriented substrates with various amounts of doped nitrogen. The orientation and coherence time of grown-in NV centers were measured in order to assess their potential use in magnetometry or other quantum mechanical applications.

## 2. Experimental details

Diamond substrates made of cylindrical (113) plates were prepared from standard 1.5 mm thick (100) HPHT *Ib* diamond crystals [28] following the procedure presented in Fig. 1. First, two parallel planes were laser-cut with an angle of 25.2° with respect to the top face, in the direction of one of the corners of the cubic-shape crystal (Fig. 1a). A cylinder having a diameter of about 2.4 mm was then laser-cut in the resulting piece (Fig. 1b). Eventually both (113) facets of the plate were polished (Fig. 1c).

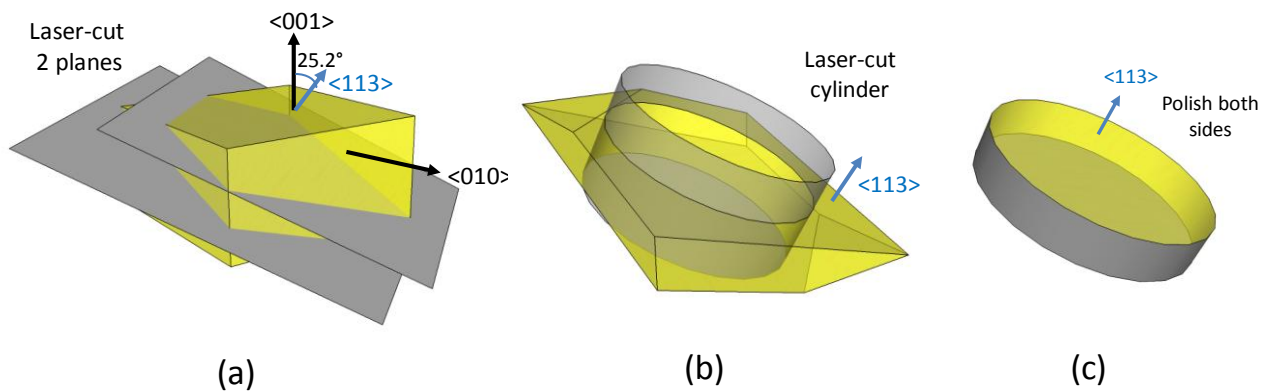

*Figure 1. Procedure for preparing cylindrical (113) plates from a (100) HPHT diamond crystal. (a) 2 parallel planes are laser-cut with a 25.2° angle, (b) a 2.4 mm-diameter cylinder is laser-cut, (c) the resulting (113) cylinder is polished on both sides.*

Plasma assisted chemical vapor deposition (PACVD) was then carried out in a home-made reactor using deposition conditions optimized for (100) growth [29]. Those include a high power density (3.5 kW, 250 mbar), a temperature of about 900 °C and a $H_2/CH_4$ gas mixture (96/4). High purity hydrogen (9N) and methane (6N) were used. A low $N_2$ amount (from 0 to 50 ppm) was intentionally added in order to tune NV density. Background contamination by nitrogen impurities is however believed to occur even when no $N_2$ is added. Growth was sustained for several hours resulting in layers having thicknesses from 460 to 1450 µm (samples A to D in Table 1). For comparison purpose, additional CVD layers were grown on (113) and (111) substrates prepared from thick optical-grade *Element 6* diamond material grown in the standard <100> direction and laser cut (samples E and F in Table 1). High-quality smooth (111)-grown layers were obtained without twinning by using a low methane concentration and a high growth temperature as explained in Ref. [22].

| Sample | Substrate used | Nitrogen doping in gas phase (ppm) | Growth Temperature (°C) | Methane concentration (%) | Thickness of CVD layer (µm) |
|---|---|---|---|---|---|
| A | (113) HPHT | 0 | ~ 900 | 4 | 460 |
| B | (113) HPHT | 0.5 | ~ 900 | 4 | 760 |
| C | (113) HPHT | 10 | ~ 900 | 4 | 500 |
| D | (113) HPHT | 50 | ~ 900 | 4 | 1450 |
| E | (113) CVD | 0 | ~ 900 | 4 | 900 |
| F | (111) CVD | 0 | ~ 1050 | 1 | 40 |

*Table 1. Growth conditions used for the CVD layers studied in this work, and thickness obtained.*

The samples were observed by Scanning Electron Microscopy (SEM) in a *ZEISS EVO MA15* system and confronted to a 3D model reported elsewhere [30] in order to identify the different faces formed during growth. Laser microscope images were acquired using a *Keyence VK 9700* equipment. Large-scale photoluminescence (PL) images were recorded with a dedicated *DiamondView*[TM] equipment which uses near band-gap UV light (around 225 nm) to excite luminescence from the samples. Raman spectra were acquired at room temperature in a *Jobin-Yvon HR800* system using the 488 nm excitation line of an Ar laser. Cathodoluminescence (CL) analysis was also performed at 110 K with a *Jobin-Yvon CLUE* system attached to the SEM on samples C, D and F as well as on a reference electronic-grade (100) CVD plate. The electron gun was operated with an accelerating voltage of 10 kV for a measured current of around 1.1 nA. The light collected by the parabolic mirror in the UV-visible range was analyzed with a *TRIAX 550* spectrometer equipped with 600 and 1800 grooves diffraction gratings. The signal was averaged when scanning an area of about 80×100 µm² (magnification of around 1000 ×).

The properties of individual NV defects hosted in (113)-oriented CVD diamond samples were analyzed with a home-made scanning confocal microscope operating under ambient conditions, as described in details elsewhere [31]. Electron spin resonance (ESR) spectroscopy was performed by applying a microwave excitation through a copper microwire directly spanned on the diamond surface. Orientation of the grown-in NV centers was assessed by recording orientation-dependent Zeeman shift of the ESR while applying a static magnetic field [16]. Finally, the spin coherence time of the NV defect electronic spin was measured by applying a spin echo sequence [$\pi/2$ - $\tau$ – $\pi$ - $\tau$ - $\pi/2$], which consists of resonant microwave $\pi/2$ and $\pi$ pulses separated by a variable free evolution duration $\tau$ [32].

### 3. CVD diamond growth on (113) substrates

A general observation is that (113)-grown layers have a smooth surface without any non-epitaxial features or twins even for thicknesses of several hundreds of micrometers (1st row of Fig. 2). Moreover, PL images recorded with the *DiamondView$^{TM}$* equipment (2nd row of Fig. 2) show only limited blue fluorescence except for the highest doping level (sample D), which suggests that residual stress remains low.

The addition of low $N_2$ amounts in the gas phase led to an increase in growth rates from 15 µm/h at 0 ppm up to around 50 µm/h at 50 ppm which is by at least a factor of 2 higher than growth rates achieved on a conventional (100) orientation under similar conditions [33]. Moreover step-bunching frequently observed in the presence of nitrogen for (100) growth [34] did not occur for (113) growth. As an indication, the surface roughness (Ra) measured for the 3 samples with lowest doping (A-C) on a 50×50 µm² was in the range 30-60 nm. However the morphology substantially degraded at the highest doping level, as illustrated in Fig. 2d.

The Raman/PL analysis of Fig. 2e shows intense and narrow diamond Raman peaks for samples A to C with a Full Width at Half Maximum (FWHM) comparable to high-quality (100)-grown layers. Sample D grown with the highest $N_2$ level shows a wider Raman peak that reaches 3.2 cm$^{-1}$ and that is shifted towards lower wavenumbers (inset of Fig. 2e). This indicates both the presence of stress and a decrease in crystalline quality. $NV^0$ and $NV^-$ peaks are detected at room temperature for samples C and D clearly indicating nitrogen incorporation in the crystals.

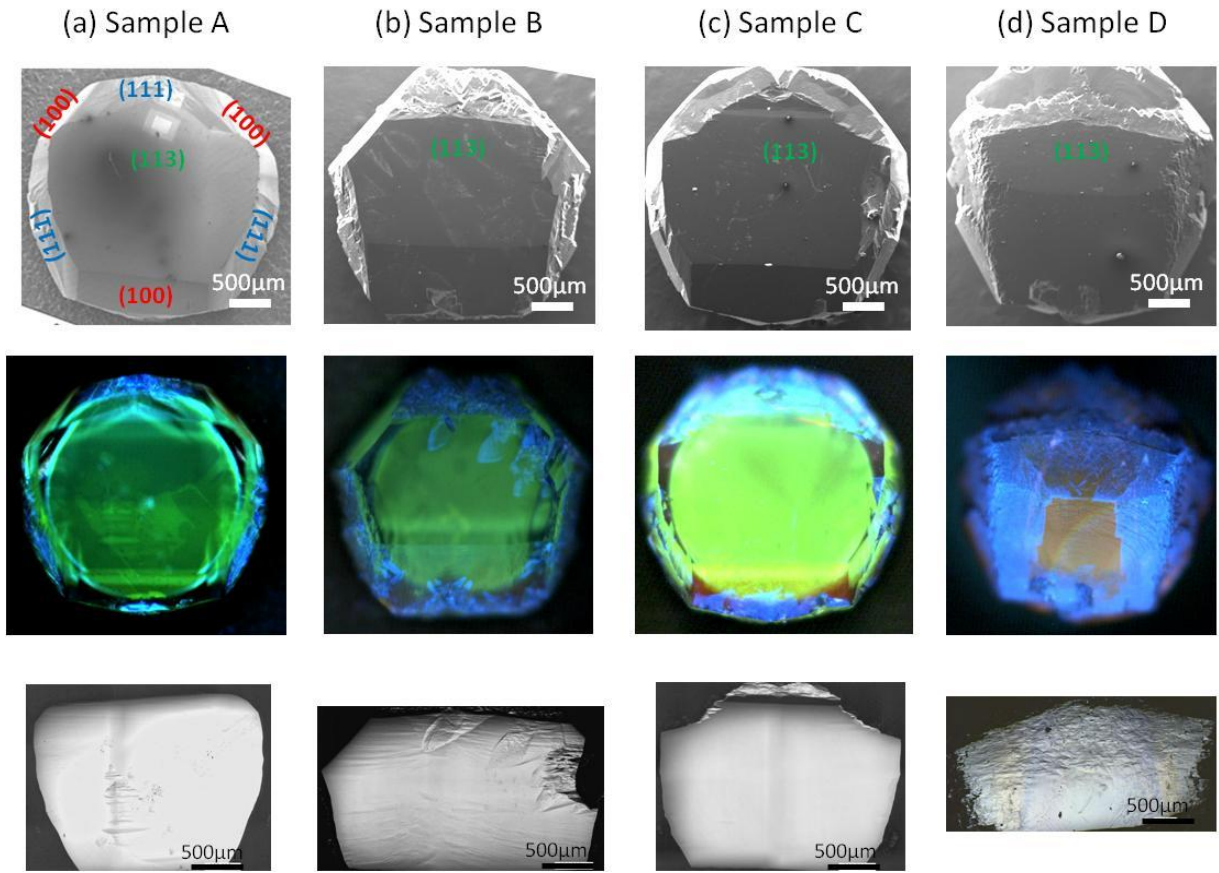

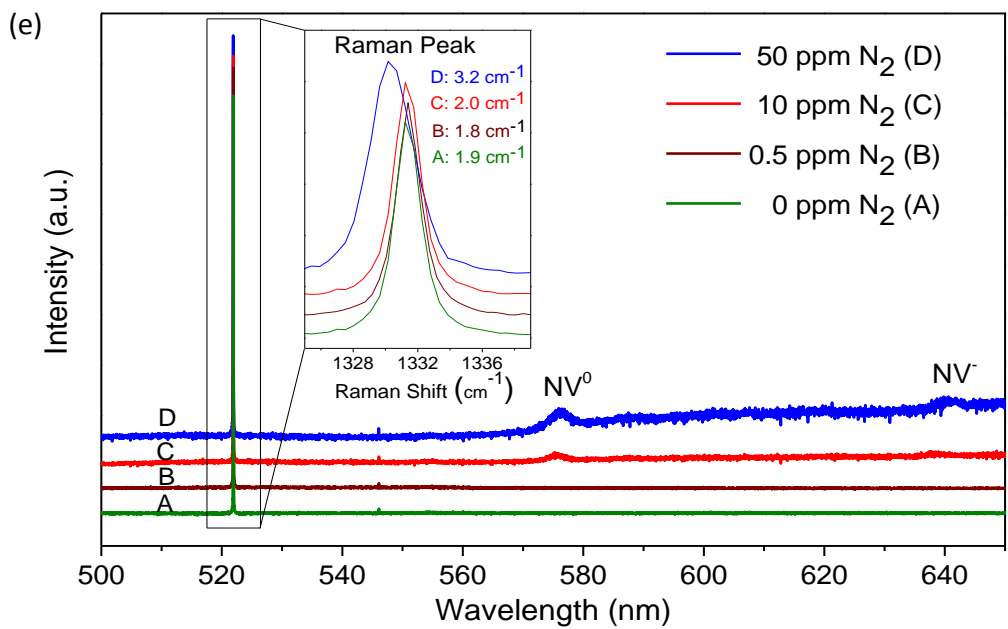

*Figure 2.* (a)-(d) CVD diamond layers grown on (113)-oriented HPHT substrates. (a) sample A, no intentional $N_2$ addition, (b) sample B, 0.5 ppm $N_2$, (c) sample C, 10 ppm $N_2$, (d) sample D, 50 ppm $N_2$ in the gas phase. First row: SEM images where the different crystalline faces can be identified; Second row: PL images recorded with the DiamondView$^{TM}$ equipment under UV light excitation. Green luminescence comes from the substrate and blue or red luminescence from stress or nitrogen related centers in the CVD layer; Third row: laser microscope mapping images of the top (113)

*face showing how the morphology is degraded at the highest doping level (d). (e) Raman/PL spectra acquired at room temperature with a 488 nm excitation line for samples A to D (from bottom to top). The spectra were normalized to the Raman peak and vertically shifted for clarity. The inset shows a zoom into the diamond Raman peak region where the FWHM of the peak can be extracted.*

Another important point that needs to be taken into account when growing thick layers on any orientation, is the appearance of additional crystalline faces that can lead to a reduction in the top surface area, or induce stress in the adjacent faces [35]. In the first row of Fig. 2a, the different crystalline faces are identified. As expected (111) faces are twinned while (100) faces are relatively smooth except when $N_2$ is added due to the well-known step-bunching phenomenon. The large (100) face that develops in the bottom part of the image tends to completely overgrow the (113) face especially when the growth thickness is high and when $N_2$ is added. Therefore a reduction in the useful top surface area is only observed for very large thicknesses (> 1 mm) or for high $N_2$ additions (> 10 ppm) indicating that this orientation is suitable to obtain large and thick high-quality layers.

To better comparatively assess crystalline quality and purity, CL analysis was carried out on samples with various crystalline orientations and doping (Fig. 3a and 3b). The CL spectrum obtained for the undoped (113) CVD layer is dominated by emission in the Free-Exciton (FE) region with only a very weak fluorescence in the visible region which is comparable to an electronic-grade (100)-grown diamond crystal (Fig. 3a). When $N_2$ is intentionally added, the contribution from $NV^0$ centers becomes obvious at 575 nm and is accompanied by several large bands in the UV-visible range which could be related to extended defects. These bands are the strongest for the smooth undoped (111)-grown layer. This indicates that despite optimized growth conditions, this orientation leads to incorporation of a larger amount of defects and stress, as previously reported [12]. In all spectra, a slight $SiV^-$ luminescence is also detected at around 737 nm which probably arises from contamination by quartz material in the reactor chamber. Nevertheless, the relatively high crystalline quality of the layers is confirmed by the presence of intense free exciton recombination assisted by both Transverse Optic (TO) and Transverse Acoustic (TA) phonons (Fig. 3b). For the (111)-grown layer, an additional peak which relates to the boron Bound Exciton recombination is observed at around 238 nm. Incorporation of residual impurities such as boron is indeed enhanced as compared to both (100) and (113) orientation. From the $BE^{TO}/FE^{TO}$ emission ratio [36], we estimate that the residual boron concentration in the film is around $8\times10^{15}$ cm$^{-3}$.

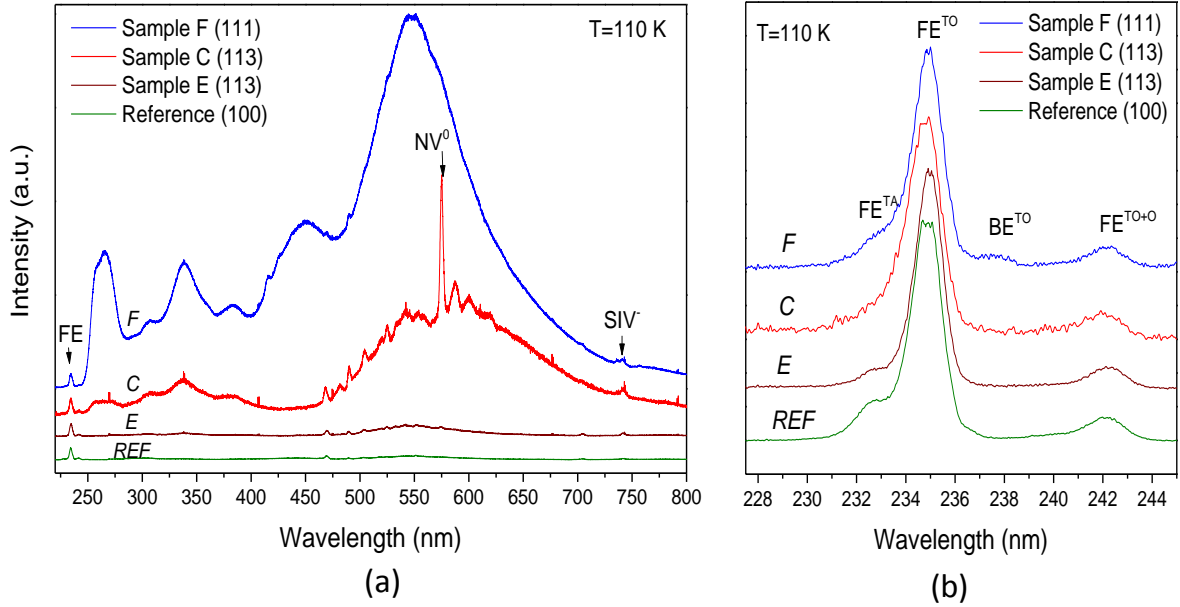

*Figure 3*. CL spectra acquired at 110 K for samples grown on different orientations and doping levels. (a) The full range spectra were normalized to the free exciton (FE) peak at around 235 nm. (b) The UV excitonic region spectra show contribution from free excitons assisted by Transverse Acoustic (TA) and Optic (TO) phonons. Boron Bound Excitons ($BE^{TO}$) are also detected for the (111)-grown layer.

In summary, (113)-grown CVD layers compare favorably to both (100) and (111). High quality layers can be obtained at high growth rates under standard high-power density growth conditions with a limited incorporation of defects and impurities. No background luminescence or stress was evidenced which is comparable to the best (100) electronic-grade material available. The surface remained smooth without step bunching while nitrogen doping could be easily tuned in the layer by adding $N_2$ in the gas phase. Degradation of crystalline quality was only observed when the added nitrogen was higher than several tens of ppm. Therefore it is believed that this material could be a useful platform for applications making use of grown-in NV color centers such as magnetometry. The ability to introduce a large amount of NV centers while preserving a good crystalline quality is particularly important when ensemble of NVs are needed since the magnetic sensitivity roughly scales with the square root of their density [2].

**4. Orientation and spin properties of NV defects in (113) layers**

We now evaluate the properties of NV centers in our (113)-grown material. To this end, individual NV defects were optically addressed in sample A using a confocal microscope under green laser excitation. A typical PL scan is shown in Fig. 4a, which indicates well-isolated spots corresponding to single NV defects. The unicity of the emitter was verified by recording the second-order autocorrelation function $g^{(2)}(\tau)$ of the PL intensity using a Hanbury Brown and Twiss interferometer. A sub-poissonian photon statistics such that $g^{(2)}(0)<0.5$ is the signature of a single emitter (Fig. 4b) [37].

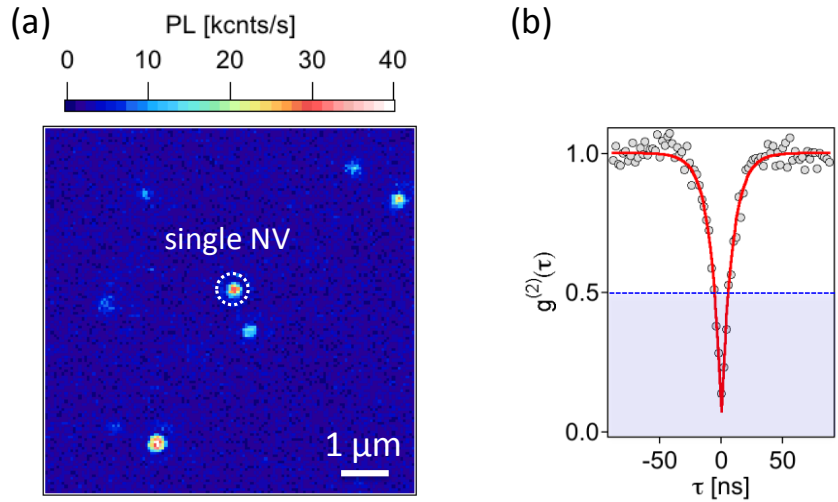

***Figure 4**. (a) PL raster scan of sample A recorded with a green laser power of 100 μW. (b) Second-order autocorrelation function $g^{(2)}(\tau)$ of the PL intensity recorded from the spot indicated with a white dashed circle in (a). A strong anti-correlation effect is observed around zero delay $g^{(2)}(0)\approx 0.1$.*

The four possible NV defect orientations are shown in Fig. 5a and the angles that they make with the <113> direction is indicated. The two orientations <1 $\bar{1}$ 1> and <$\bar{1}$ 1 1> make an equal angle α=58.5° with the <113> direction. The <1 1 $\bar{1}$> orientation is almost in the (113) plane (α=100°) while the <111> direction points out with an angle α=29.5°. The NV defect orientation was then experimentally measured by recording ESR spectra while applying a static magnetic field $B$= 18 G perpendicular to the (113) surface plane (i.e. along the <113> direction). In this limit of weak magnetic fields [2], the ESR frequencies are given by $\nu_\pm = D \pm g\mu_B B_{NV}$, where $D$=2.87 GHz is the zero-field splitting of the NV defect, $g\mu_B \sim 28\ GHz/T$ and $B_{NV} = B \times \cos\alpha$ is the magnetic field projection along the NV defect quantization axis. For each NV defect, measurement of the ESR spectrum therefore enables to discriminate between NV defect with <111> orientation (α=29.5°), {<1 $\bar{1}$ 1> ; <$\bar{1}$ 1 1>} orientations (α=58.5°) and <1 1 $\bar{1}$> orientation (α=100°), as illustrated in Fig. 5b. The probability of occurrence of each NV defect orientations was estimated by recording ESR spectra over a set of about 200 single NV defects. The resulting statistical distribution is shown in Fig. 5d. We observe a significant preferential orientation of NV defects along the <111> direction with a probability of about 73 %. It is striking that centers that are almost in the (113) plane (centers oriented along <1 1 $\bar{1}$>) were not detected. This is consistent with previously reported results on NV orientation dependence on crystal growth direction, which usually showed that "in-plane" directions are very unfavorable for the creation of NVs [13-17]. On the other hand, directions closest to the growth direction are promoted, such as <111> in this case. Although the precise mechanism leading to preferential alignment of NV centers in (113) layers cannot be fully determined here, models taking into account the atomistic configuration and the movement of steps during growth could be proposed in the future, such as that reported for (111)-grown layers [38].

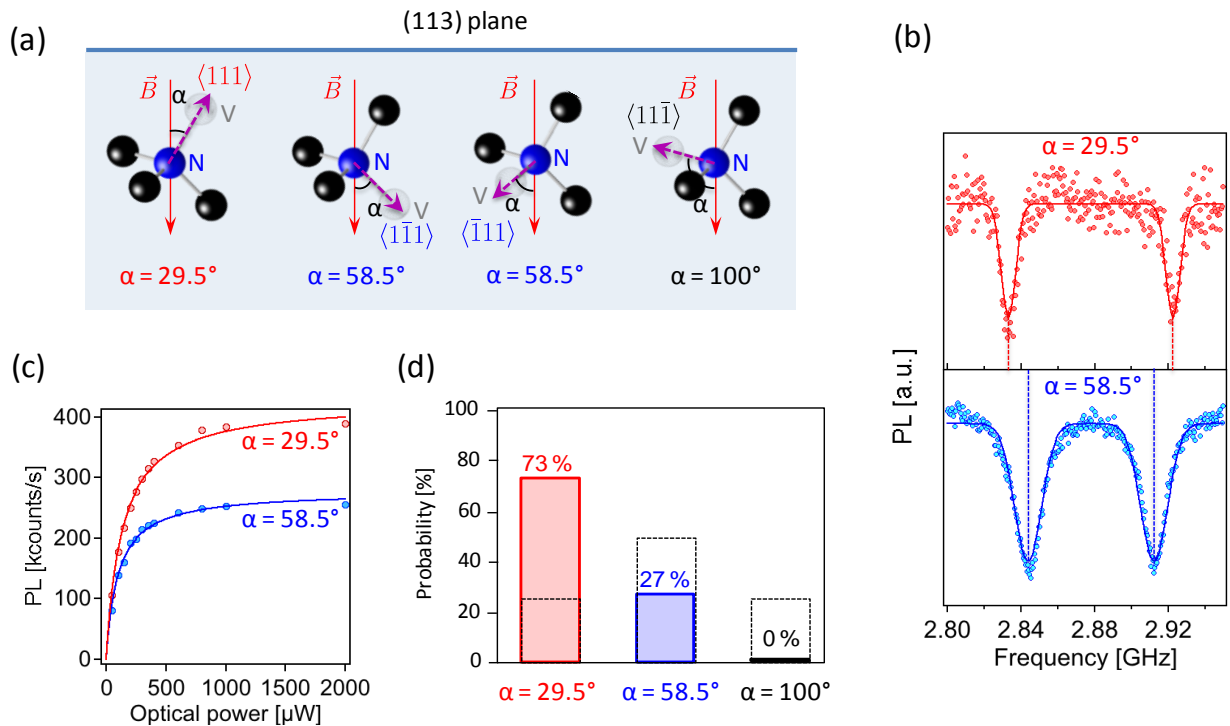

*Figure 5. (a) Schematic drawing of the four possible orientations of NV defects with respect to the <113> direction. (b) Orientation-dependent ESR spectra recorded from single NV defects in sample A while applying a static magnetic field B=18 G perpendicular to the (113) diamond surface plane (c) Collected PL intensity as a function of the laser power for single NV defects with different orientations. (d) Statistical distribution of NV defect orientations extracted from ESR measurements for a set of about 200 single NV defects. The black dashed lines indicate the expected distribution for randomly oriented NV defects.*

We note that the collected PL intensity is significantly larger for NV defects oriented along the preferential <111> direction (Fig. 5c), which is an additional advantage. This is a direct consequence of the direction of the NV defect optical dipoles, which are lying in a plane perpendicular to the NV axis [16]. As a consequence from the small α angle for the <111> direction, light extraction efficiency from the (113) plane is thus enhanced.

It is also relevant to evaluate the spin coherence properties of NV centers in our (113)-grown material and compare to those measured for centers incorporated during growth on a conventional (100) orientation. A typical spin-echo measurement recorded from a single NV defect in sample A is depicted in Fig. 6. It shows collapses and revivals of the electron spin coherence induced by Larmor precession of $^{13}C$ nuclear spin impurities [32]. The envelope of the spin echo signal indicates a coherence time $T_2$ of ≈270 μs. This is comparable to a single NV defect hosted in standard high-purity (100) CVD crystals with the same isotopic purity.

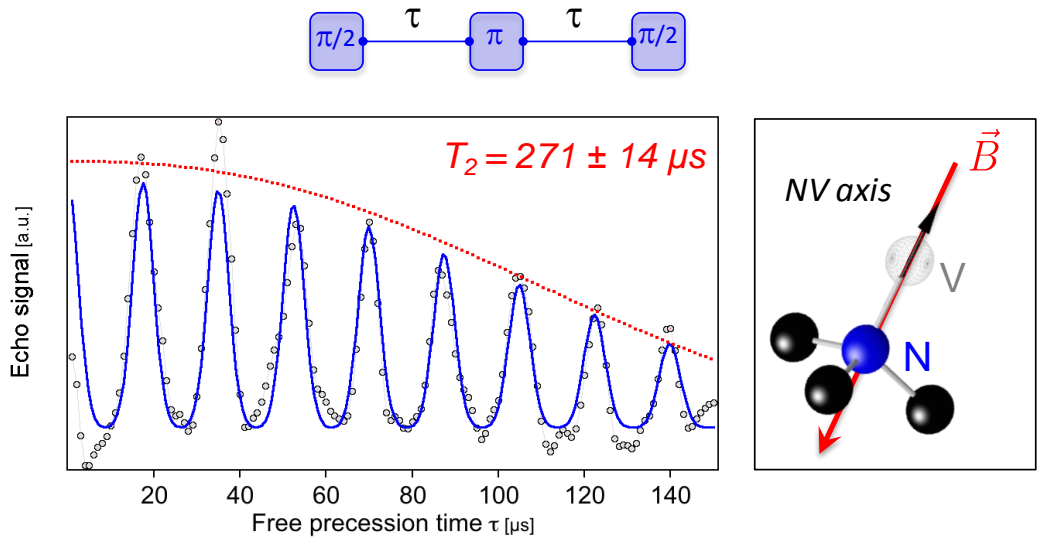

*Figure 6. Spin echo signal recorded from a grown-in <111>-oriented single NV defect in a (113) oriented CVD layer. A magnetic field B=53 G is applied along the NV defect axis.*

**Conclusion**

In this work, diamond layers were epitaxially grown by PACVD on (113) substrates prepared from thick HPHT and CVD crystals by laser cutting and polishing. Standard high-power conditions resulted in growth rates in the range 15-50 μm/h. Smooth morphologies without step bunching were observed for several hundreds of μm thick films. The crystalline quality of undoped layers was comparable to that of electronic-grade commercial (100) material and substantially superior to that of smooth layers (111)-grown layers. Tuning of NV center density could be achieved in a wide range by intentionally adding nitrogen to the gas phase.

Analysis of NV centers in (113)-grown layers has revealed a preferential orientation in the <111> direction with a probability of about 73 %; the remaining 27 % being shared between two other equivalent orientations: <1 $\bar{1}$ 1> and <$\bar{1}$ 1 1>. The most in-plane direction (<1 1 $\bar{1}$>) was not detected. This is consistent with previous reports that highlighted that out-of-plane directions are more favorable for NV centers. Finally NV centers in (113)-grown layers exhibited coherence times as high as that measured in conventional (100) crystals with a similar isotopic purity.

Although preferential orientation is only partial as compared to the recently reported results for (111)-grown layers, the ease of growth and processing on (113) substrates opens the way to their use in the fabrication of useful structures for magnetometry or other quantum mechanical applications.

**Acknowledgements:**

The authors would like to thank A. Edmonds and M. Markham from *Element 6* (UK) for providing some of the (113) and (111) CVD diamond substrates used in this study. This research has been supported by the European Community's Seventh Framework Programme (FP7/2007-2013) under Grant Agreement n◦ 611143 (DIADEMS) and by the French Agence Nationale de la Recherche through the project ADVICE (ANR-2011-BS04-021). Cathodoluminescence set-up was partially funded by C'Nano Ile de France (SMECAL AAP-2011) and the French Labex SEAM (Science and Engineering of Advanced Materials).